**RESEARCH**

**Open Access**

# Few-shot genes selection: subset of PAM50 genes for breast cancer subtypes classification

Leandro Y. S. Okimoto[1*], Rayol Mendonca-Neto[1], Fabíola G. Nakamura[1], Eduardo F. Nakamura[1], David Fenyö[2] and Claudio T. Silva[2]

*Correspondence:
okimoto@icomp.ufam.edu.br

[1] Institute of Computing, Universidade Federal do Amazonas, Manaus, BR, Brazil
[2] New York Univesity, New York, USA

**Abstract**

**Background:** In recent years, researchers have made significant strides in understanding the heterogeneity of breast cancer and its various subtypes. However, the wealth of genomic and proteomic data available today necessitates efficient frameworks, instruments, and computational tools for meaningful analysis. Despite its success as a prognostic tool, the PAM50 gene signature's reliance on many genes presents challenges in terms of cost and complexity. Consequently, there is a need for more efficient methods to classify breast cancer subtypes using a reduced gene set accurately.

**Results:** This study explores the potential of achieving precise breast cancer subtype categorization using a reduced gene set derived from the PAM50 gene signature. By employing a "Few-Shot Genes Selection" method, we randomly select smaller subsets from PAM50 and evaluate their performance using metrics and a linear model, specifically the Support Vector Machine (SVM) classifier. In addition, we aim to assess whether a more compact gene set can maintain performance while simplifying the classification process. Our findings demonstrate that certain reduced gene subsets can perform comparable or superior to the full PAM50 gene signature.

**Conclusions:** The identified gene subsets, with 36 genes, have the potential to contribute to the development of more cost-effective and streamlined diagnostic tools in breast cancer research and clinical settings.

## Background

Advances in omic techniques have revolutionized how we analyze gene expression data, enabling us to understand various diseases, including cancer. One such disease, breast cancer, is known for its heterogeneity, with different subtypes exhibiting distinct biological characteristics and treatment responses [1–4]. In recent years, a wealth of genomic and proteomic data has become available, offering valuable insights into the underlying biology of breast cancer. However, harnessing this amount of information requires efficient theoretical frameworks, instruments, and computational tools [5, 6].

Breast cancer subtypes, such as hormone receptor-positive and hormone receptor-negative, can be further classified into more specific subgroups based on their intrinsic properties. For example, hormone receptor-positive breast cancer can be further classified into luminal A, luminal B, or HER2-enriched subtypes, while hormone receptor-negative breast cancer can be classified into triple-negative or basal-like subtypes [7]. In recent years, the PAM50 gene signature, a set of 50 genes used to classify breast cancer into the four intrinsic molecular subtypes, has emerged as a valuable prognostic tool in breast cancer research, providing insights into tumor subtypes and guiding therapeutic decision-making [8]. However, the PAM50 approach has some limitations, including its reliance on a relatively large number of genes, which can increase costs and complexity in research and clinical settings [9]. This highlights the need for a more efficient method to classify breast cancer subtypes using fewer genes accurately.

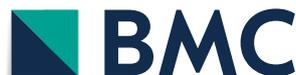

The primary objective of this research is to explore the potential for achieving precise breast cancer subtype categorization utilizing a reduced gene set derived from the PAM50 gene signature. By employing a method known as "Few-Shot Gene Selection", we will randomly select smaller subsets from PAM50 and evaluate their performance using the F1 Score and a linear model, specifically, the Support Vector Machine (SVM) classifier. In addition, this approach assesses whether a more compact gene set can maintain accuracy while simplifying classification.

Our contributions are as follows:

1. An experiment demonstrating the impact of using fewer genes in the gene selection phase;
2. A direct comparison with the PAM50 gene selection method;
3. A 2D visualization of t-distributed stochastic neighbor embedding (t-SNE) applied to the model's predicted samples, providing insight into how the model perceives the data, including misclassified samples at subtype boundaries.

Through this research, we hope to demonstrate that selecting fewer but more representative genes can lead to a more accurate classification of breast cancer subtypes while addressing the limitations of the PAM50 approach.

## Related works

In this section, we discuss the related work on the limitations of the PAM50 gene signature in the context of breast cancer subtype classification. Also, we briefly present some works related to Gene Expression Data Selection.

### PAM50 gene signature

The PAM50 gene signature, also referred as the Prosigna Breast Cancer Prognostic Gene Signature Assay, is a molecular classification method that assists in identifying breast cancer subtypes [10]. This gene signature utilizes the expression levels of 50 specific genes (Table 1) to classify breast cancer into four intrinsic subtypes: Luminal A, Luminal B, HER2-enriched, and Basal-like. The accurate classification of these subtypes is critical, as they are associated with different prognoses and treatment options [11].

In clinical practice, the PAM50 gene signature is employed as a prognostic tool to predict the likelihood of breast cancer recurrence, guide treatment decisions, and estimate patient survival [8]. It is particularly useful for determining the most appropriate treatment for patients with early-stage, hormone receptor-positive breast cancer. It can help identify those at a higher risk of recurrence who may benefit from more aggressive treatment options.

**Table 1** PAM50 Gene Signature

| Gene Set | Genes |
| --- | --- |
| PAM50 | ACTR3B, ANLN, BAG1, BCL2, BIRC5, BLVRA, CCNB1, CCNE1, CDC20, CDC6 NUF2, CDH3, CENPF, CEP55, CXXC5, EGFR, ERBB2, ESR1, EXO1, FGFR4, FOXA1, FOXC1, GPR160, GRB7, KIF2C NDC80, KRT14, KRT17, KRT5, MAPT, MDM2, MELK, MIA, MKI67, MLPH, MMP11, MYBL2, MYC, NAT1, ORC6, PGR, PHGDH, PTTG1, RRM2, SFRP1, SLC39A6, TMEM45B, TYMS, UBE2C, UBE2T |

The prognostic value of the PAM50 gene signature in predicting improved outcomes has shown some limitations [12, 13]. One of these limitations is using unsupervised statistical methods in deriving the PAM50 signature, which may result in the inclusion of genes that lack biological relevance to breast cancer prognosis or treatment [14]. Consequently, the PAM50 gene signature might miss biologically meaningful genes crucial for understanding breast cancer prognosis and treatment [13].

Additionally, some studies using the PAM50 gene signature apply arbitrary cut-off values when evaluating the functional groups of genes [9]. This method may further contribute to the exclusion of important genes. Adopting a more nuanced approach in selecting cut-off values and incorporating more relevant genes could potentially improve the performance and clinical utility of the PAM50 gene signature [9, 14].

**Gene selection in expression data**

Mendonca-Neto et al. [15] proposed a novel outlier-based gene selection (OGS) strategy to pick significant genes for classifying breast cancer subtypes quickly and effectively. In a test set of 77 samples, the authors present a multi-level classifier demonstrating that their technique yields $F1$ scores of 1.0 for basal and 0.86 for her 2, the subtypes with the poorest prognoses, respectively. The authors suggested strategy surpasses existing methods in terms of $F1$ score, using 80% fewer genes. Their strategy generally picks just a few highly important genes applied to a hierarchical classifier, accelerating classification and enhancing performance considerably.

Liu et al. [16] proposed a gene selection method combining double radial basis function (RBF) kernels with weighted analysis to extract relevant genes from gene expression data. By eliminating redundant and irrelevant genes, this method addresses the challenge of analyzing gene expression data with many genes and small samples. The authors tested the modified method on four benchmark datasets, including two-class and multiclass phenotypes. They found it outperformed previous methods regarding accuracy, true positive rate, false positive rate, and reduced runtime. This approach allows for knowledge-based interpretation of omics data, providing essential information about various biological processes and reflecting the current physiological status of cells and tissues.

Yang et al. [17] developed a hybrid approach that combines correlation-based feature selection and binary particle swarm optimization to select relevant gene subsets from microarray gene expression data for disease classification and medical diagnosis. They applied this approach to six gene expression datasets related to human cancer. They used the K-nearest neighbor method as a classifier to evaluate classification performance. The results demonstrated that the proposed approach effectively simplified the feature selection process and reduced the number of parameters needed while achieving higher classification accuracy than other feature selection methods. Their method could be an ideal pre-processing tool for optimizing the feature selection process, improving classification accuracy while minimizing computational resources. Furthermore, this approach may also apply to other problem areas.

**Our approach**

While these studies have contributed significantly to gene selection and breast cancer classification, there is still room for improvement. One aspect often overlooked in these studies is the importance of comparing the proposed methods with the current state-of-the-art, such as the PAM50 gene signature. By comparing novel gene selection techniques with the PAM50, we can better assess their true potential and practical relevance in real-world clinical and research settings.

In our experiments, we aim to address this gap by comparing the performance of our proposed gene selection strategy with the well-established PAM50 gene signature. This comparison will help us understand the advantages and limitations of our approach and evaluate its potential to provide accurate and efficient breast cancer subtype classification in real-world situations. Furthermore, it will enable us to identify areas where further improvements can be made, ultimately leading to more robust and effective gene selection methods for breast cancer subtype categorization.

**Approach explained**

In this study, we introduce an innovative approach termed "Fewer-Shot Genes Selection" to refine breast cancer subtype classification. Our method strategically generates multiple gene subsets stemming from the well-established PAM50 signature, subsequently evaluating their potency in classification. One of the crucial motivations behind this approach is the profound practical significance of reducing gene sets. In medicine, a condensed set of genes simplifies the classification process. Clinicians and researchers no longer need to identify a vast array of genes, making the diagnostic procedure more efficient and potentially more accurate. By critically assessing the performance of these gene subsets, our goal is to unearth combinations that either match or surpass the classification prowess of the PAM50 signature, thus enhancing the overall process of breast cancer subtype classification. The exhaustive pipeline of our proposed method is illustrated in Fig. 1.

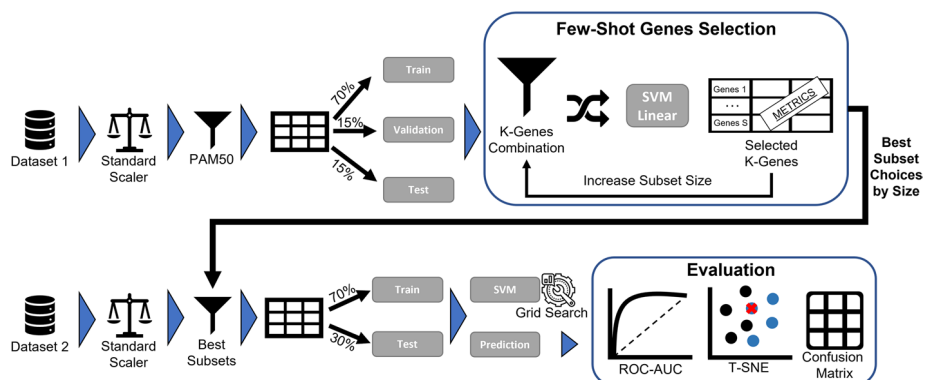

**Fig. 1** Proposed approach

In summary, our approach consists of the following steps:

1. We take one dataset and normalize it using the Standard Scaler.
2. We filter the dataset using the PAM50 signature.
3. We split the filtered dataset into the train, validation, and test sets, ensuring the Gene Selection will not suffer overfitting.
4. We input the train, validation, and test sets into the "Few-Shot Genes Selection" to assess the optimal subset choices based on the size of the Gene Signature.
5. We use the best subset choices in a second pipeline, which starts by normalizing a second dataset using the Standard Scaler.
6. We filter the second dataset using each best subset choice (Gene Signatures from different sizes).
7. We evaluate each filtered subset using a train/test Support Vector Machine (SVM) with Grid Search. Here we do not need a validation portion, once there is no data leak from the gene selection phase from Dataset 1.
8. We evaluate the test predictions using the t-Distributed Stochastic Neighbor Embedding (t-SNE), Confusion Matrix, and Receiver Operating Characteristic (ROC) - Area Under the Curve (AUC) curve.

This comprehensive approach, splitting Dataset 1 into *Train, Validation, and Test* to ensure unbiased subset selection, and Dataset 2 only into *Train and Test*, for the evaluation process, enables the selection of optimal gene subsets and their subsequent evaluation without bias from any Dataset.

**Few-shot genes selection**

In this study, we adopted a 'Few-shot gene selection' approach, wherein subsets ranging from 10% to 80% of the total signature length were analyzed. This strategy was carefully chosen to optimize the balance between model accuracy and interpretability, in addition to reducing computational demands. Specifically, for the PAM50 Gene Signature, this approach entails evaluating subsets from size 5 to size 40. To achieve this, we used a Support Vector Machine (SVM) with a linear kernel, which was trained on 70% of the data, validated on 15%, and tested on the remaining 15%.

In the rigorous evaluation of gene subsets within the "Few-shot gene selection" methodology, we executed a comprehensive set of experiments, conducting 1 million trials for each combination of subset size and experimental setup. Given the vastness of potential combinations, assessing each exhaustively is infeasible. Thus, we strategically organized our evaluation into three separate experiments, each encompassing 1 million distinct combinations, resulting in an aggregate of 3 million combinations per subset size. Post-experimentation, a statistical analysis was performed to compare the outcomes of the three experiments. The results consistently registered a p-value greater than 0.05, indicating that the differences in the experiments were not statistically significant. This affirms the robustness of our methodology, ensuring that our observations are not attributable to random variations.

This extensive experimentation allowed us to explore a large search space and identify the optimal gene subsets that maximize the combined $F$1 scores from both validation and test data for the ACES and TCGA datasets separately. After this validation, we combined the data from all three experiments for each subset size, selected the gene sets corresponding to the maximum validation and tested combined $F$1 scores. We then cross-evaluated the best choices for one dataset on the other dataset and vice versa, ensuring that each subset size achieved the results without overfitting from one dataset.

**Evaluation**

Our study evaluated the performance of the "Few-Shot Genes Selection" approach by comparing the results with the well-established PAM50 gene signature. In addition, we employed several evaluation metrics and visualization techniques to highlight the differences between the PAM50 and the best-chosen subset.

1. ROC-AUC Curve: We used the Receiver Operating Characteristic (ROC) curve and calculated the Area Under the Curve (AUC) for PAM50 and our top 3 subset size results. This enables us to compare the performance of the gene subsets in terms of their sensitivity and specificity, providing a comprehensive view in comparison with PAM50 Signature.
2. Confusion Matrix: We generated confusion matrices for the PAM50 gene signature and the subset size results to assess the classification accuracy, allowing us to identify any discrepancies in the classification performance between PAM50 and our selected gene subsets.
3. T-SNE: We employed t-Distributed Stochastic Neighbor Embedding (t-SNE) to visualize the high-dimensional gene expression data in a lower-dimensional space. By comparing the t-SNE plots of the PAM50 gene signature with the results of our method, we could identify patterns and structures in the data and analyze any differences in the grouping of samples.

Using these evaluation methods, we can effectively compare the performance of our "Few-Shot Genes Selection" approach with the PAM50 gene signature.

## Data and methods

In this section, we describe the datasets used in our experiments. We also present the preprocessing and machine learning algorithm employed in our paper, and also the evaluation metrics applied.

### Experiment datasets

A large dataset is required to classify breast cancer subtypes without overfitting accurately [18]. To do this, samples from 12 ACES studies (n = 1606) [19] and the TCGA breast invasive cancer dataset (n = 532) [20] were used. These datasets, covering 13 independent investigations, form a collection including 2138 samples. Only the METABRIC dataset [21], which needs ethical approval, is absent from the data presented in this research, making our study the most extensive compilation of public gene expression data for Breast Cancer subtypes.

Our datasets should capture a substantial proportion of the biological heterogeneity across breast cancer patients and the technical biases resulting from the variation in platforms and study-specific sample preparations [22]. This heterogeneity will aid the trained models in achieving a higher degree of generalization, which is vital for applying the final classification model in the real world [23–25]. In this study, the inclusion criteria were specified to concentrate on particular molecular subtypes of breast cancer. The original dataset contained 2,148 samples. For our analysis, we included samples classified under the subtypes: **Basal**, **HER2**, **Luminal B**, and **Luminal A**, while excluding samples identified as Normal-like and those without defined subtypes. This led to a refined dataset comprising 2027 samples, bringing **ACES to 1512** and for **TCGA to 515** samples.

The focus of our research is on the post-surgical treatment of patients diagnosed with breast cancer. A key component of this research involves the classification of tumors into subtypes using small gene signatures. This approach is essential for determining appropriate treatment modalities. The inclusion of normal samples is not relevant to our study's objectives, as our emphasis is on treatment strategies for specific tumor subtypes, not on early detection or the transformation from normal tissue to cancer. Additional information about clinical variables was added to the Additional file 1.

### Experiment settings and computation time

We conducted our experiments on a custom server equipped with a 12th Gen Intel(R) Core(TM) i9-12900KF processor, boasting 16 cores. The system was complemented with 64 GB of RAM and a 1 TB NVMe SSD for storage. All computational tasks were performed on an environment running Ubuntu 20.04.6 LTS. Regarding computation time, the Experiment with TCGA as a filter required approximately 48 h, while the Experiment ACES took close to 52 h to access the list of best subsets.

We used Python programming language and scikit-learn[1] package for the machine learning algorithms to perform all the experiments in this manuscript.

---

[1] https://scikit-learn.org/stable/

**Data preprocessing**

*PAM50 gene selection*

For gene filtering, we applied the PAM50 Gene Selection [8]. The PAM50 signature was produced using unsupervised statistical approaches; hence, no restrictions were placed on the identified genes' biological significance beyond their prediction ability. Unanswered is the issue of which regulatory factors are responsible for the diverse expression patterns of this group of genes in the distinct molecular subtypes [26]. It is one of the most used gene filters in the literature, which is the one we are going to use as the basis for our analysis. In this study, we explore the potential of achieving precise breast cancer subtype categorization using a reduced gene set derived from the PAM50 gene signature.

*Standard scaler*

Standard Scaler is used when the ranges of the input dataset's attributes differ significantly or when they are measured in various units of measure. As we deal with 13 distinct breast cancer studies, we must normalize our data by removing the mean and scaling to the unit variance.

Standardization:

$$z = \frac{x - \mu}{\sigma} \quad (1)$$

Here, $z$ represents the standardized value, $x$ is the original value of a data point, $\mu$ is the mean of the dataset, and $\sigma$ is the standard deviation of the dataset.

With mean:

$$\mu = \frac{1}{N} \sum_{i=1}^{N} (x_i) \quad (2)$$

In this equation, $\mu$ is the mean of the dataset, $N$ is the number of data points, and $x_i$ represents each individual data point in the dataset.

And standard deviation:

$$\sigma = \sqrt{\frac{1}{N} \sum_{i=1}^{N} (x_i - \mu)^2} \quad (3)$$

Here, $\sigma$ represents the standard deviation of the dataset, $N$ is the number of data points, $x_i$ is each individual data point in the dataset, and $\mu$ is the mean of the dataset.

Utilizing the Standard Scaler, we adjust our data to a distribution where the mean is normalized to 0 and the standard deviation to 1. Specifically, for our multivariate dataset, this adjustment occurs on a feature-by-feature basis, meaning it is applied independently to each feature across all samples. This approach is particularly vital in the context of our study, which leverages machine learning techniques.

Given the sensitivity of machine learning algorithms to the distribution and scale of input data, the Standard Scaler plays a crucial role in minimizing the possibility of biases or distortions. These might otherwise stem from batch effects, thereby ensuring that each sample is processed on its own merits. This method is essential for maintaining the accuracy and reliability of our machine learning models, contrasting with techniques like quantile normalization and RMA. While the latter adjust data against a common reference distribution, they may not be as suited for machine learning contexts where batch effects could significantly impact results, potentially leading to biased outcomes in machine learning analyses.

**SVM**

We decided to utilize traditional machine learning instead of employing deep learning. We know how deep learning methods have advanced in the cancer area through some works [27–29]. Still, as our dataset size and filtering method lead us to few samples for training and high-dimension features with gene expression, a Deep Learning algorithm could suffer from overfitting and high-variance gradient updates [30].

Support vector machines (SVM) in machine learning are supervised learning models with associated learning algorithms that evaluate data for regression and classification. SVMs, based on statistical learning frameworks, are among the most reliable prediction techniques [31]. SVMs may effectively do non-linear classification in addition to linear classification by implicitly translating their inputs into high-dimensional feature spaces. This technique is known as the kernel trick. The choice of the SVM was based on a previous study where the SVM method achieved the best results for representative genes [32].

**Evaluation**

Here we discuss the evaluation methods used to assess the performance of our model. We focus on two key aspects: the ROC-AUC curve, the t-SNE visualization, and various performance metrics, including accuracy, precision, recall, F1 score, and AUC.

*ROC-AUC curve*

The Receiver Operating Characteristic (ROC) curve is a graphical representation of the true positive rate (sensitivity) versus the false positive rate (1-specificity) for different classification thresholds. The Area Under the Curve (AUC) of the ROC curve is a scalar value ranging from 0 to 1, which measures the overall performance of a classifier. For example, a model with perfect classification would have an AUC of 1. In contrast, a random classifier would have an AUC of 0.5. The ROC-AUC curve is a useful tool for evaluating and comparing the performance of different classifiers.

*Metrics*

To further evaluate the performance of our model, we compute several metrics: accuracy, precision, recall, F1 score, and AUC.

- **Accuracy**: The proportion of correctly classified instances out of the total number of instances. The formula for accuracy is:

$$Accuracy = \frac{TP + TN}{TP + TN + FP + FN} \tag{4}$$

where TP, TN, FP, and FN represent true positives, true negatives, false positives, and false negatives, respectively.

- **Precision**: The proportion of true positive instances among the instances predicted as positive. The formula for precision is:

$$Precision = \frac{TP}{TP + FP} \tag{5}$$

- **Recall** (Sensitivity): The proportion of true positive instances among the actual positive instances. The formula for recall is:

$$Recall = \frac{TP}{TP + FN} \tag{6}$$

- **F1 Score**: The harmonic mean of precision and recall, which provides a balanced measure of both metrics. The formula for the F1 score is:

$$F1\ Score = 2 \times \frac{Precision \times Recall}{Precision + Recall} \tag{7}$$

- **AUC**: As mentioned earlier, the Area Under the Curve (AUC) of the ROC curve is a scalar value that measures the overall performance of a classifier. A higher AUC value indicates better classifier performance.

*Visualization: t-SNE*

We used a dimensionality reduction technique known as a t-distributed stochastic neighbor embedding (t-SNE) to show the results of our studies obtained from our experiment. t-SNE is a popular dimensionality reduction technique for visualizing high-dimensional data in lower-dimensional spaces, such as two or three dimensions. the main objective of t-SNE is to preserve the data structure by maintaining the pairwise relationships between data points when projecting them onto a lower-dimensional space [33].

The t-SNE algorithm involves three main steps:

1. Compute pairwise affinities $p_{ij}$ in the high-dimensional space using a Gaussian distribution:

$$p_{ij} = \frac{\exp\left(-||x_i - x_j||^2 / 2\sigma^2\right)}{\sum_{k \neq l} \exp\left(-||x_k - x_l||^2 / 2\sigma^2\right)} \tag{8}$$

2. Compute pairwise affinities $q_{ij}$ in the low-dimensional space using a Student's t-distribution with one degree of freedom:

$$q_{ij} = \frac{(1 + ||y_i - y_j||^2)^{-1}}{\sum_{k \neq l}(1 + ||y_k - y_l||^2)^{-1}} \tag{9}$$

3. Minimize the Kullback–Leibler divergence (KL divergence) [34] between the two distributions concerning the positions of the points in the map using the following cost function:

**Table 2** SVM parameters used in our experiments. Parameters missing were set as default values in scikit-learn version 1.4.0

| Parameter | Value |
| --- | --- |
| C | 1 |
| Kernel | Linear |
| Gamma | Scale |
| Decision function shape | One-vs-rest |

$$C = \sum_{i \neq j} p_{ij} \log \frac{p_{ij}}{q_{ij}} \qquad (10)$$

In these formulas, $x_i$ and $x_j$ are data points in the high-dimensional space, while $y_i$ and $y_j$ are data points in the low-dimensional space. The variable $\sigma$ represents the variance of the Gaussian distribution in the high-dimensional space, and it is usually chosen using a binary search for each data point. The pairwise probabilities $p_{ij}$ and $q_{ij}$ represent the likelihood of data points $i$ and $j$ being similar in the high-dimensional and low-dimensional spaces, respectively.

By minimizing the KL divergence (cost function $C$), t-SNE ensures that the relationships between data points are preserved when projecting the data onto a lower-dimensional space. This makes t-SNE an effective method for visualizing complex datasets and uncovering hidden structures within the data.

This algorithm is broadly used among researchers in cancer with gene expression data [35–37]. It can represent samples in data points scattered on a plot and give us insights into the subtypes' relationship.

## Results

In this section, we examine the findings of our cross-validation approach. This technique involves determining the optimal subset for one dataset and subsequently applying it to the model on the second dataset, then reversing the process to select from the second dataset and applying it to the first dataset. Our aim is to capture less complex attributes in the Selection Phase using a linear model while employing a Grid Search to train non-linear SVM models in the Evaluation Phase, the parameters applied in the SVM models are shown in Table 2.

In the Data Preprocessing step we had to do a gene intersection between datasets ACES and TCGA, in order to select one dataset and ensure that we had the genes in the other dataset. When we applied the PAM50 gene in the intersection, we found that the ACES dataset did not have 6 genes presented in the PAM50 Gene Signature: *ANLN, CXXC5, GPR160, NUF2, TMEM45B, UBE2T*. This does not discourage us from using the PAM50 gene signature because the list still had 44 important genes to use as features in our models.

### TCGA to ACES

Initially, we utilized the TCGA dataset with 515 samples as the filtering dataset. After passing this dataset to the first pipeline with the "Few-Shot Genes Selection" process, we retained the best subsets for each subset length then we applied the ACES dataset for the evaluation process.

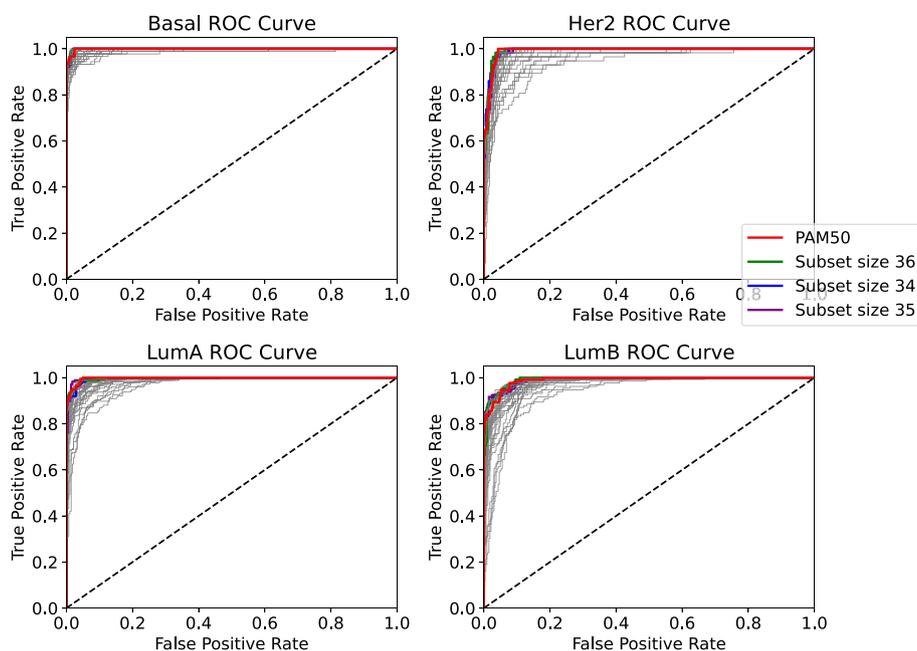

**Fig. 2** ROC curve for the evaluation of the prediction samples from ACES in a model with features filtered by TCGA

*ROC-AUC*

In the ROC curve, we highlighted the top 3 subsets which achieved AUC (Area Under Curve) equal to or even higher than the baseline PAM50, highlighted in red in Fig. 2, Subset size 36 (S-36), Subset size 35 (S-35) and Subset size 34 (S-34).

As we can see, there are selections that can compete with the PAM50 Gene Signature baseline, the gray lines in the plot also show us other selections which contain subsets with fewer genes.

*Metrics*

The results in Table 3 present the performance of four highlighted filtering methods (PAM50, S-36, S-35, and S-34) across the four subtypes (Basal, Her2, LumA, and LumB), reporting Accuracy (Acc), Precision (Prec), Recall (Rec), F1 Score (F1), and AUC for each case.

In the Basal subtype, S-35 attained the highest accuracy (0.943), while S-36 reached the best F1 Score (0.961). PAM50, however, secured the highest precision (0.966) and AUC (0.999), with both S-36 and S-35 achieving the best recall (0.978). For Her2, S-35 displayed the highest accuracy (0.943) but lower precision (0.852) and F1 Score (0.829) compared to PAM50 and S-36. Conversely, PAM50 and S-34 tied for the highest AUC (0.992), and PAM50 demonstrated the best recall (0.860), with S-34 leading in precision (0.882).

Table 3 Classification results for the four subtypes TCGA to ACES

|           | Basal |       |       |       |       | Her2  |       |       |       |       |
|-----------|-------|-------|-------|-------|-------|-------|-------|-------|-------|-------|
|           | Acc   | Prec  | Rec   | F1    | AUC   | Acc   | Prec  | Rec   | F1    | AUC   |
| PAM50     | 0.930 | **0.966** | 0.944 | 0.955 | **0.999** | 0.930 | 0.845 | **0.860** | **0.852** | **0.992** |
| Subset 36 | 0.927 | 0.946 | **0.978** | **0.961** | **0.999** | 0.927 | 0.855 | 0.825 | 0.839 | **0.992** |
| Subset 35 | **0.943** | 0.935 | **0.978** | 0.956 | **0.999** | **0.943** | 0.852 | 0.807 | 0.829 | 0.991 |
| Subset 34 | 0.927 | 0.955 | 0.955 | 0.955 | **0.999** | 0.927 | **0.882** | 0.789 | 0.833 | **0.992** |
|           | LumA  |       |       |       |       | LumB  |       |       |       |       |
|           | Acc   | Prec  | Rec   | F1    | AUC   | Acc   | Prec  | Rec   | F1    | AUC   |
| PAM50     | 0.930 | 0.955 | 0.966 | 0.960 | 0.997 | 0.930 | 0.908 | 0.902 | 0.905 | 0.990 |
| Subset 36 | 0.927 | 0.954 | 0.949 | 0.952 | 0.997 | 0.927 | 0.909 | 0.909 | 0.909 | **0.992** |
| Subset 35 | **0.943** | **0.967** | **0.989** | **0.978** | **0.998** | **0.943** | **0.953** | 0.917 | **0.934** | 0.991 |
| Subset 34 | 0.927 | 0.944 | 0.960 | 0.952 | 0.996 | 0.927 | 0.904 | **0.924** | 0.914 | 0.991 |

In LumA, S-35 outperformed other methods with the highest accuracy (0.943), precision (0.967), recall (0.989), F1 Score (0.978), and AUC (0.998). For LumB, S-35 achieved the highest accuracy (0.943) and F1 Score (0.934), and S-36 obtained the highest AUC (0.992). Additionally, S-35 had the highest precision (0.953), and S-34 presented the best recall (0.924).

Overall, S-35 consistently performed well across all subtypes, obtaining the highest accuracy and F1 Score in most situations. Nevertheless, PAM50 exhibited better precision for Basal and the highest metrics for Her2. Comparing the results, S-35 can be considered the best-performing method in most cases, with S-36 and PAM50 remaining competitive in some aspects.

*Confusion matrix and t-SNE*

In the Confusion Matrix, we highlight the subset size of 36, which achieved the highest mean AUC for the four subtypes.

Comparing both Confusion Matrices, Fig. 3 and 4, we can observe that both had the same hits for Basal and LumB subtypes, while the PAM50 achieved better results in LumA and our method achieved better results in Her2.

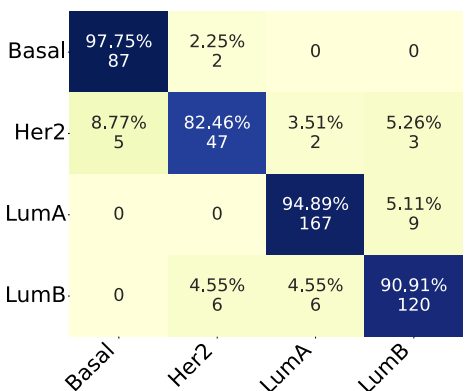

**Fig. 3** Confusion Matrix for subset size 36, filtering from TCGA and applying on ACES

|       | Basal | Her2 | LumA | LumB |
|-------|-------|------|------|------|
| Basal | 97.75%<br>87 | 2.25%<br>2 | 0 | 0 |
| Her2  | 10.53%<br>6 | 80.7%<br>46 | 0 | 8.77%<br>5 |
| LumA  | 0 | 0 | 98.3%<br>173 | 1.7%<br>3 |
| LumB  | 0 | 3.03%<br>4 | 6.06%<br>8 | 90.91%<br>120 |

**Fig. 4** Confusion Matrix for PAM50 applying on ACES

Upon examining both t-SNE visualizations, Fig. 5 and 6, we can see that both had almost the same clear separation between the subtypes. We highlight the errors of each plot that could be identified on the Confusion Matrices. Of course, it's not a perfect separation once we're trying to project n-dimensional features in a 2D Visualization. Nevertheless, we can gain an understanding of how the data perceives in our model and which samples are missing in the boundaries of each subtype.

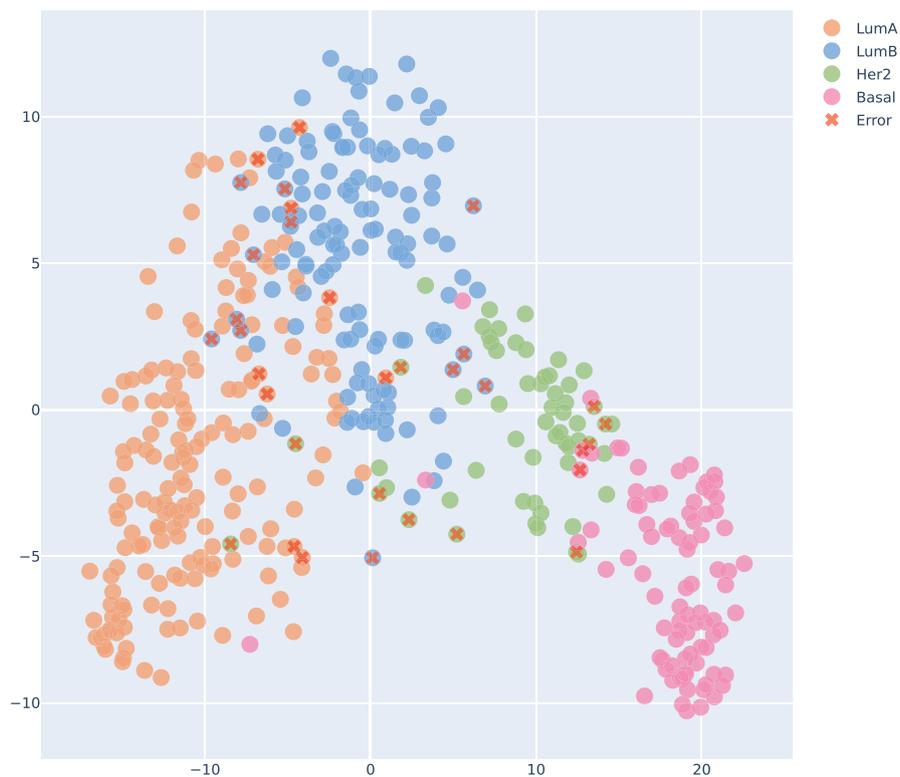

**Fig. 5** t-SNE Visualization for subset size 36, filtering from TCGA and applying on ACES

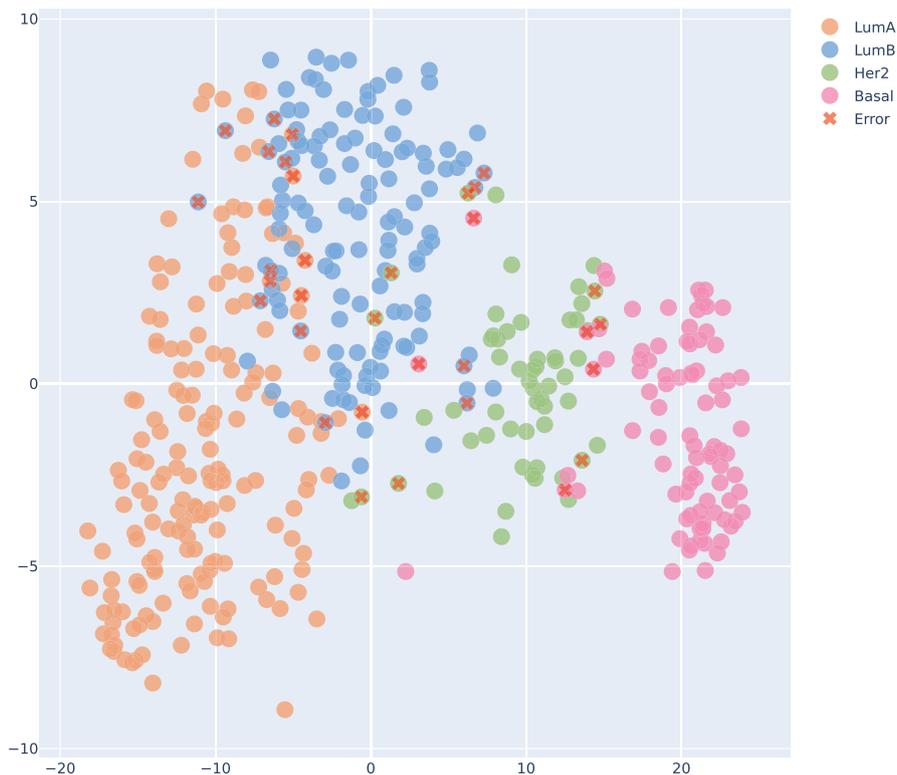

**Fig. 6** t-SNE Visualization in the ACES Prediction data after PAM50 filtering

**ACES to TCGA**

After doing the TCGA filtering and ACES dataset evaluation, we cross-evaluated, using the ACES dataset with 1512 samples as the filtering dataset. After choosing the best subsets for each subset length, we applied the TCGA dataset for the evaluation process. Our goal by doing so is to have a full understanding of how fewer genes impact the model's performance when we compare it against the PAM50 Signature.

*ROC-AUC*

In the ROC curve, we highlighted the top 3 subsets which achieved AUC (Area Under Curve) equal to or even higher than the baseline PAM50, highlighted in red in Fig. 7, Subset size 37 (S-37), Subset size 36 (S-36) and Subset size 32 (S-32).

As we can see, the results achieved by the subsets can compete with the PAM50 baseline. In this particular ROC-AUC visualization, where we have less data in the prediction with the TCGA dataset, we can have a better view of the subsets that achieved greater curves than the baseline.

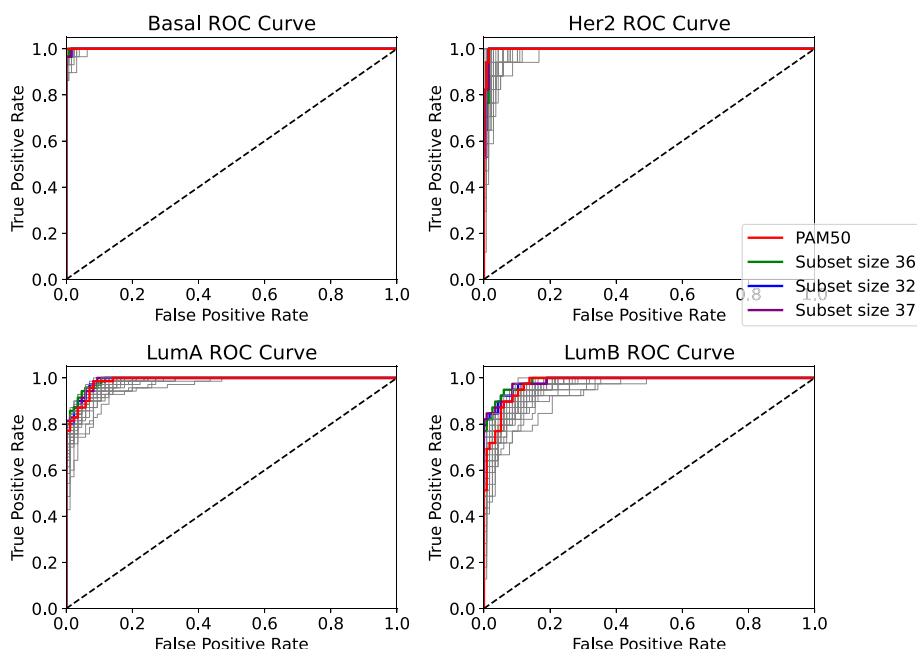

**Fig. 7** ROC curve for the evaluation of the prediction samples from TCGA in a model with features filtered by ACES

*Metrics*

Table 4 presents the performance of four filtering methods (PAM50, S-37, S-36, and S-32) across four subtypes (Basal, Her2, LumA, and LumB), reporting Accuracy (Acc), Precision (Prec), Recall (Rec), F1 Score (F1), and AUC for each case.

In the Basal subtype, S-36 achieved the highest accuracy (0.935), while PAM50, S-36, and S-32 shared the best precision (1.000), recall (0.931), F1 Score (0.964), and AUC (1.000). For Her2, S-36 displayed the highest accuracy (0.935) and shared the best recall (1.000) with PAM50 and S-32. S-32 attained the highest precision (0.895) and F1 Score (0.944), as well as the highest AUC (0.997).

**Table 4** Classification results for the four subtypes ACES to TCGA

|  | Basal | | | | | Her2 | | | | |
| --- | --- | --- | --- | --- | --- | --- | --- | --- | --- | --- |
|  | Acc | Prec | Rec | F1 | AUC | Acc | Prec | Rec | F1 | AUC |
| PAM50 | 0.910 | **1.000** | **0.931** | **0.964** | **1.000** | 0.910 | 0.850 | **1.000** | 0.919 | 0.999 |
| Subset 37 | 0.910 | **1.000** | 0.897 | 0.945 | 0.999 | 0.910 | 0.842 | 0.941 | 0.889 | 0.996 |
| Subset 36 | **0.935** | **1.000** | **0.931** | **0.964** | **1.000** | **0.935** | 0.850 | **1.000** | 0.919 | 0.996 |
| Subset 32 | 0.923 | **1.000** | **0.931** | **0.964** | 0.999 | 0.923 | **0.895** | **1.000** | **0.944** | **0.997** |
|  | LumA | | | | | LumB | | | | |
|  | Acc | Prec | Rec | F1 | AUC | Acc | Prec | Rec | F1 | AUC |
| PAM50 | 0.910 | 0.914 | 0.914 | 0.914 | 0.988 | 0.910 | 0.868 | 0.846 | 0.857 | 0.978 |
| Subset 37 | 0.910 | 0.915 | 0.929 | 0.922 | 0.990 | 0.910 | 0.872 | **0.872** | 0.872 | 0.985 |
| Subset 36 | **0.935** | **0.944** | **0.957** | **0.950** | **0.992** | **0.935** | **0.919** | **0.872** | **0.895** | **0.988** |
| Subset 32 | 0.923 | 0.929 | 0.929 | 0.929 | 0.989 | 0.923 | 0.872 | **0.872** | 0.872 | 0.986 |

In LumA, S-36 outperformed other methods with the highest accuracy (0.935), precision (0.944), recall (0.957), F1 Score (0.950), and AUC (0.992). For LumB, S-36 secured the highest accuracy (0.935), precision (0.919), F1 Score (0.895), and AUC (0.988), while sharing the best recall (0.872) with S-37 and S-32.

Overall, S-36 consistently performed well across all subtypes, achieving the highest accuracy, F1 Score, and AUC in most cases, and sharing the best results in precision and recall with other methods. S-32 also demonstrated competitive performance in several aspects, particularly for the Her2 subtype.

*Confusion matrix and t-SNE*

In the Confusion Matrix, by coincidence, we highlight the subset size of 36, which achieved the highest mean AUC for the four subtypes.

Comparing both Confusion Matrices, Fig. 8 and 9, we can observe that both had the same hits for Basal and Her2 subtypes. But our subset size 36 had better performance on LumA and LumB subtypes.

If we take a look at both t-SNE visualizations, Fig. 10 and 11, we can see that both had almost the same clear separation between the subtypes, the difference was an inversion of the y axis but that was not because of our features, it was just the non-linear

|       | Basal | Her2 | LumA | LumB |
|-------|-------|------|------|------|
| Basal | 93.1% / 27 | 6.9% / 2 | 0 | 0 |
| Her2  | 0 | 100.0% / 17 | 0 | 0 |
| LumA  | 0 | 0 | 95.71% / 67 | 4.29% / 3 |
| LumB  | 0 | 2.56% / 1 | 10.26% / 4 | 87.18% / 34 |

**Fig. 8** Confusion Matrix for subset size 36, filtering from ACES and applying on TCGA

|       | Basal | Her2 | LumA | LumB |
|-------|-------|------|------|------|
| Basal | 93.1% / 27 | 6.9% / 2 | 0 | 0 |
| Her2  | 0 | 100.0% / 17 | 0 | 0 |
| LumA  | 0 | 1.43% / 1 | 91.43% / 64 | 7.14% / 5 |
| LumB  | 0 | 0 | 15.38% / 6 | 84.62% / 33 |

**Fig. 9** Confusion Matrix for PAM50 applying on TCGA

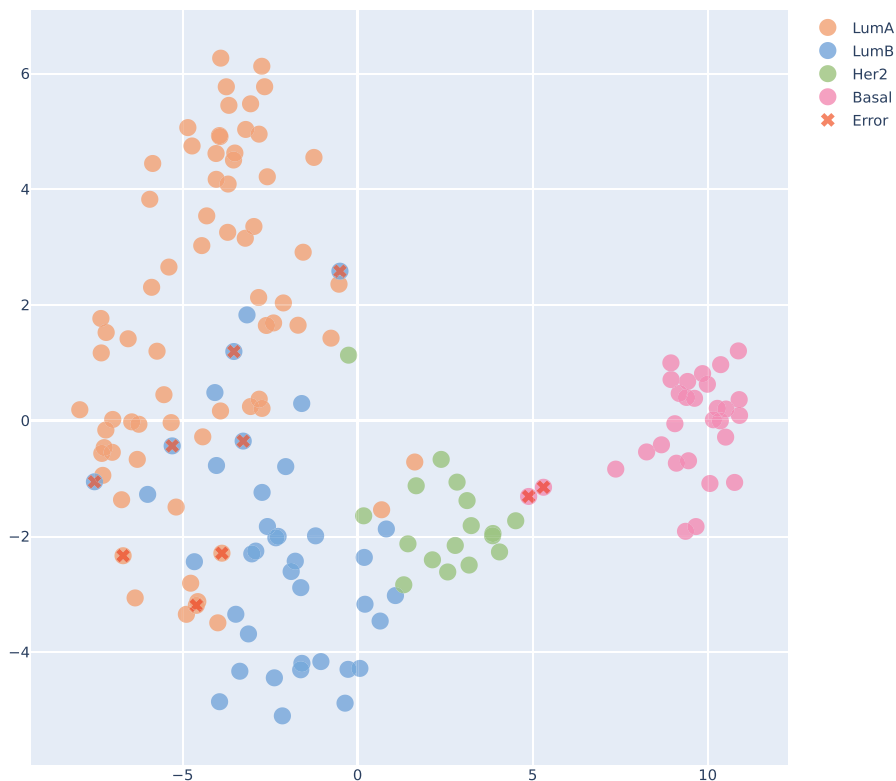

**Fig. 10** t-SNE Visualization subset size 36, filtering from ACES and applying on TCGA

transformation of t-SNE method. We highlight the errors of each plot that could be identified on the Confusion Matrices. We can clearly see the errors in the subtypes' boundaries.

**Discussion**

In our study, we observed that the list of genes from the S-36 filtering method differed between the two datasets. Interestingly, 30 genes were common between the gene sets obtained for both datasets, shown in Table 5, suggesting that these shared genes play a significant role in cancer subtype classification. The bold entries in the table represent genes that do not intersect between the two gene sets. This finding further supports the robustness of the S-36 filtering method as it maintains its performance across different datasets, despite the variations in the gene lists.

The enhanced performance of the S-36 method compared to the PAM50 Signature across various evaluation metrics suggests that certain genes within the PAM50 Signature might not be as crucial or indicative for classifying cancer subtypes. This finding prompts a reevaluation of the PAM50 Signature's gene components due to concerns about its effectiveness. Further analysis of molecular functions using the Panther Classification System [38] has revealed that while the S-36 Signature from TCGA shares a similar distribution of molecular functions with PAM50, the S-36

TCGA - with PAM50 Signature

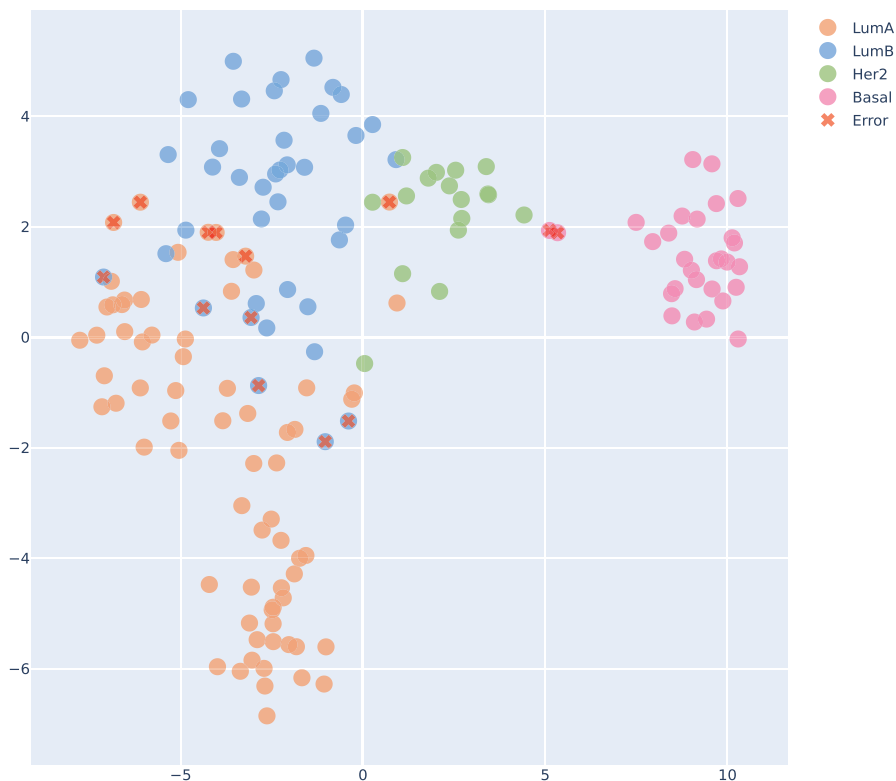

**Fig. 11** t-SNE Visualization in the ACES Prediction data after PAM50 filtering

**Table 5** S-36 Signatures for ACES and TCGA datasets experiments

| Gene Set | Genes |
|---|---|
| S-36 ACES Dataset | MAPT, BLVRA, CENPF, GRB7, BCL2, FGFR4, CDC6, CCNB1, FOXC1, KRT17, SFRP1, PTTG1, FOXA1, NDC80, KIF2C, MELK, MMP11, CEP55, SLC39A6, KRT14, MIA, NAT1, EGFR, EXO1, BIRC5, ACTR3B, CDH3, ESR1, MLPH, MDM2 **CDC20, MYBL2, PGR, MYC, PHGDH, RRM2** |
| S-36 TCGA Dataset | MAPT, BLVRA, CENPF, GRB7, BCL2, FGFR4, CDC6, CCNB1, FOXC1, KRT17, SFRP1, PTTG1, FOXA1, NDC80, KIF2C, MELK, MMP11, CEP55, SLC39A6, KRT14, MIA, NAT1, EGFR, EXO1, BIRC5, ACTR3B, CDH3, ESR1, MLPH, MDM2 **UBE2C, CCNE1, MKI67, TYMS, KRT5, BAG1** |

Signature from ACES didn't include the "Structural Molecular Activity" provided by the KRT5 gene, which is unique in this aspect. Additional information about gene functions was added to the Additional file 2.

These results indicate potential areas for enhancing the PAM50 Signature. By improving the process of gene selection, it is conceivable to create more precise and effective classifiers for cancer subtypes. The common genes identified by the S-36 filtering methods in both datasets could be a valuable focus for future research, as they may be critical to improving cancer subtype classification accuracy. Consequently, this could offer avenues to classify Breast Cancer subtypes using a more concise gene signature.

In summary, our study not only highlights the potential of the S-36 filtering method as a viable alternative to the PAM50 Signature, but also brings attention to the need for reevaluating the PAM50 Signature gene components. By focusing on the shared genes and improving the gene selection process, we can potentially develop more accurate and efficient classifiers for cancer subtypes, ultimately contributing to better patient care and management.

## Conclusion and future directions

In this paper, we developed an approach called "Fewer-Shot Genes Selection" which utilized a large number of subsets derived from the PAM50 Signature, in order to compare the classification performance of this baseline.

We achieved results as S-36 that surpassed the baseline PAM50 Signature, which indicates that there is potential for improvement in the feature selection techniques for a better Breast Cancer Subtype classification.

The results achieved in this paper encourage us to extend our approach to a higher dimension, not limiting ourselves to a subset of the PAM50 Signature, but the entire feature space. As some papers suggest in the background showing the limitations of the PAM50 Signature, we could prove that even by taking subsets derived from PAM50, we could compare favorably with it.

We also consider applying explainable artificial intelligence (XAI) in the evaluation process directly in the misclassified samples in the boundaries between subtypes, which could be explained by the expression of specific features or even a group of features for that error, this explainable AI process could be really valuable for future methods where we could take into account of how certain features contribute in the misclassification of some samples that are in the subtype boundaries.

### Supplementary Information

The online version contains supplementary material available at https://doi.org/10.1186/s12859-024-05715-8.

**Additional file 1.** Additional Clinical Variables.

**Additional file 2**. Genes and functions.

**Funding**
This study was financed in part by the Coordenação de Aperfeiçoamento de Pessoal de Nível Superior - Brasil (CAPES-PROEX) - Finance Code 001. This work was partially supported by Amazonas State Research Support Foundation - FAPEAM - through the POSGRAD project.

**Availability of data and materials**
The data that support the findings of this study are openly available in GitHub at https://github.com/UMCU-Genetics/SyNet/tree/master/Gene_Expression_Datasets. Given the extensive number of samples analyzed in our study, we have included a comprehensive list of the PatientIDs, which serve as identifiers for both the TCGA and ACES datasets, within a Additional file. This Additional file 1 is named "AdditionalClinicalVariables.xlsx".

## Declarations

**Ethics approval and consent to participate**
No ethics approval was required for this study.

**Consent for publication**
Not applicable.

**Competing interests**
The authors declare that they have no competing interests.

Received: 6 December 2023   Accepted: 21 February 2024
Published online: 01 March 2024

**Publisher's Note**